\documentclass[a4paper,11pt]{article}
\usepackage{pos}
\usepackage{subcaption}
\usepackage{physics}

\title{\vspace{-2cm}
	\begin{flushright}
		{\normalsize INR-TH-2024-004}
	\end{flushright}
	\vspace{0.5cm} Dark photon emission in elastic proton bremsstrahlung}

\author*[a, b]{Ekaterina Kriukova}

\affiliation[a]{Institute for Nuclear Research of the Russian Academy of Sciences,\\
60th October Anniversary pr-ct 7a, Moscow, Russia}

\affiliation[b]{Faculty of Physics, Lomonosov Moscow State University,\\
Leninskiye Gory 1-2, Moscow, Russia}

\emailAdd{kryukova.ea15@physics.msu.ru}

\abstract{We study the production of dark photons with masses in range 0.4-1.5\,GeV in elastic proton bremsstrahlung. In contrast to the known approaches to calculating the bremsstrahlung cross section, we consider the non-zero momentum transfer between protons. The numerical estimates show that the refined result agrees well with the generalized Weizsacker-Williams approximation, although it significantly differs from the result obtained by Blumlein and Brunner \cite{Blumlein:2013cua} which is widely used in proton beam-dump phenomenology.}

\FullConference{International Conference on Particle Physics and Cosmology (ICPPCRubakov2023)\\
 02-07, October 2023\\
Yerevan, Armenia\\}

\renewcommand{\logo}{\relax}

\begin{document}
\maketitle

\section{Dark photons}
One of the most popular ways to describe the structure of any extension of the Standard Model (SM) of particle physics is to consider it as the coexistence of the SM particles, the so-called portal and the hidden sector (HS). The portal is the way to write down the interaction between the SM and the HS. It is realized by new physics particles called mediators that can interact both with the particles from the SM sector and the hidden one. 

There are only three renormalizable portals --- scalar, vector and fermion ones \cite{Lanfranchi:2020crw}. Scalar portal is produced by the cubic and quartic interaction of dark scalars with the SM Higgs doublet. The fermion portal includes heavy neutral leptons as mediators that couple to SM leptons and the Higgs doublet via Yukawa-like interaction. Finally, the vector portal, that we consider in this paper, consists of the dark photons which are the massive vector particles that kinetically mix with the SM photons.

In this work we consider the so-called minimal dark photon model (BC1 in \textit{Beyond Colliders at CERN} classification \cite{Beacham:2019nyx}), where dark photons interact only with the SM particles and any interaction of dark photons with the dark matter is neglected. Thus one can write part of the Lagrangian relevant for our study as \cite{Okun:1982xi}
	\begin{equation}
		\mathcal{L}=\mathcal{L}_\text{SM}-\frac{1}{4}F'_{\mu \nu}F'^{\mu \nu}+\frac{\epsilon}{2}F'_{\mu\nu}B^{\mu \nu}+\frac{m_{\gamma'}^2}{2}A'_\mu A'^{\mu},
	\end{equation}
where $\mathcal{L}_\text{SM}$ is the SM Lagrangian, $F'_{\mu \nu}$ is the dark photon strength tensor, $B_{\mu \nu}$ is the SM $U(1)_Y$ strength tensor. Then the model is characterized by two parameters --- the kinetic mixing parameter $\epsilon$ and the dark photon mass $m_{\gamma^\prime}$.

Among the experiments that have performed searches for minimal dark photons (see \cite{Agrawal:2021dbo} for review) there are NA64 \cite{NA64:2018lsq}, PS191 \cite{Gninenko:2011uv}, BaBar \cite{BaBar:2014zli}, LHCb \cite{LHCb:2019vmc}, CMS \cite{CMS:2020krr}, etc. It is important to note that despite being intensively studied, the minimal dark photon parameter plane $(\epsilon, m_{\gamma^\prime})$ is still weakly constrained for $m_{\gamma^\prime}>0.8\text{ GeV}$. It will be possible to study the phenomenology of $\mathcal{O}(1)$ GeV dark photons produced in $pp$-collisions in the near detectors of neutrino experiments being prepared such as DUNE \cite{DUNE:2020fgq} and T2K \cite{deNiverville:2016rqh}. The proposed SHiP experiment at CERN could also search for visible decays of dark photon \cite{SHiP:2020vbd}.

\begin{figure}[h]
	\begin{center}
		\includegraphics[width=\textwidth]{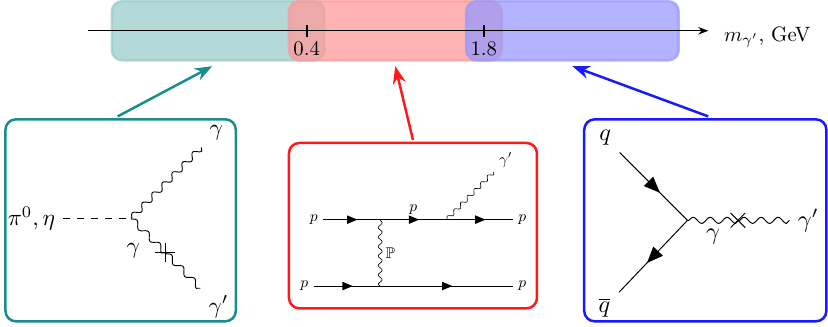}
		\caption{The dominant mechanisms of dark photon production in $pp$-collisions depending on the dark photon mass $m_{\gamma^\prime}$ from left to right: meson decays, proton bremsstrahlung and Drell-Yan process.}
		\label{fig:production}
	\end{center}
\end{figure}

Figure~\ref{fig:production}
depicts three possible mechanisms of dark photon production in $pp$-collisions. Which mechanism prevails is determined by the dark photon mass $m_{\gamma^\prime}$ \cite{Miller:2021ycl}. Dark photons lighter than 0.4 GeV are mostly produced via $\pi^0$, $\eta$ meson decays as shown in green in figure~\ref{fig:production}. The analogue of QCD Drell-Yan process shown in blue is dominant for heavier dark photons with masses above about 1.8 GeV. Red boxes illustrate the process of proton bremsstrahlung that is crucial for dark photon masses in range 0.4-1.8 GeV and was studied in this work.

The paper is organized as follows. Section \ref{sec:approximations} reviews the previously known estimates of bremsstrahlung cross section. In section \ref{sec:calculation} we present the result of our calculation of elastic proton bremsstrahlung taking into account the non-zero momentum transfer between protons. The numerical results and comparison with other methods are summarized in section \ref{sec:numerical}. Finally, section \ref{sec:conclusions} contains conclusions.

\section{Previously suggested approximations for proton bremsstrahlung} \label{sec:approximations}
\begin{figure}[b]
	\begin{center}
		\begin{subfigure}{0.4\textwidth}
			\centering
                \includegraphics[width=\textwidth]{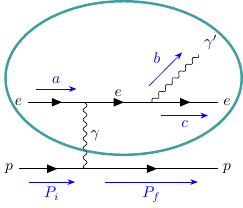}
			\caption{}
			\label{fig:WWscheme}
		\end{subfigure}
		\hfill
		\begin{subfigure}{0.4\textwidth}
			\centering
			\includegraphics[width=\textwidth]{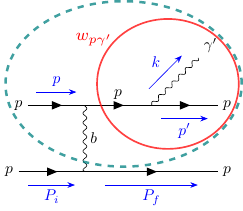}
			\caption{}
			\label{fig:BBscheme}
		\end{subfigure}
	\end{center}
	\caption{Feynman diagrams schematically illustrating the known approximations: (a) the Weizsacker-Williams approximation for electron bremsstrahlung \cite{Kim:1973he}, (b) the Blumlein-Brunner approximation for proton bremsstrahlung \cite{Blumlein:2013cua}. The subdiagram relevant to splitting function $w_{p\gamma^\prime}$ is shown in red.}
\end{figure}

Figure~\ref{fig:WWscheme}
shows the idea of generalized Wiezsacker-Williams (WW) approximation for the process of electron bremsstrahlung $e(a)p(P_i)\rightarrow \gamma^\prime(b)e(c)p(P_f)$ with one-photon exchange \cite{Kim:1973he}. Since the impact of photon propagator to the cross section is proportional to ${1}/{t^2}$, where $t\equiv-q^2$ is the minus square of photon momentum $q$ transferred between protons, the largest contribution is given at the minimal square of momentum transfer $t_\text{min}$,
\begin{equation}
     \sqrt{t_\text{min}}=\frac{(a \cdot b)-m^2_{\gamma^\prime}/2}{a_0-b_0}.
\end{equation}
This allows one to connect the answer with the cross section of the $2\rightarrow 2$ subprocess $e\gamma \rightarrow \gamma^\prime e$ that is highlighted with green ellipse in figure~\ref{fig:WWscheme},
\begin{equation}
    \left[ \frac{\dd^2 \sigma(ep\rightarrow \gamma^\prime ep)}{\dd (a\cdot b) \dd (b\cdot P_i)} \right]_{WW} = \left[\frac{\dd \sigma(e\gamma \rightarrow \gamma^\prime e)}{\dd (a\cdot b)}\right]_{t=t_{min}}\frac{\alpha}{\pi}\frac{\chi}{(c \cdot P_i)},
\end{equation}
where due to the smallness of $t_\text{min}$ the photon is usually taken on-shell, $\alpha$ is the electromagnetic coupling constant and $\chi$ is the flux of photons emitted by the target proton
\begin{equation}
    \chi=\int_{t_\text{min}}^{t_\text{max}}\frac{t-t_\text{min}}{t^2}G_2(t) \dd t.
\end{equation}
Here $G_2(t)$ is the proton electric form factor (for details see \cite{Kim:1973he}). The generalized WW approximation has been used earlier to calculate the dark photon production in electron \cite{Bjorken:2009mm} and elastic proton \cite{Foroughi-Abari:2021zbm} bremsstrahlung.

Another widely used approximation schematically shown in figure~\ref{fig:BBscheme} was formulated by Blumlein and Brunner for inelastic bremsstrahlung \cite{Blumlein:2013cua}, but can be applied to the elastic process $p(p)p(P_i)\rightarrow \gamma^\prime(k) p(p^\prime)p(P_f)$ along the same lines. In the BB approach the $pp$-interaction is considered as the exchange of hypothetical massless vector particles $b$. On the first step in the framework of WW approximation the matrix element of proton bremsstrahlung is connected to the amplitude of the $2\rightarrow 2$ subprocess $pb\rightarrow \gamma^\prime p$. Then one puts the 4-momentum of $b$-boson exactly to 0 and associates the probability of dark photon emission in the subprocess $p\rightarrow \gamma^\prime p$ with the splitting function $w_{p\gamma^\prime}(z, k_\perp^2)$ in the style of DGLAP QED equation. The first argument of the splitting function $z\equiv k_z/p_z$ is the ratio of the longitudinal component of dark photon's 3-momentum to the same component of incident proton's 3-momentum in the lab frame. The second argument $k_\perp^2$ is the square of the transverse component of dark photon's 3-momentum. The final result for cross section is
\begin{equation}
    \left[ \frac{\dd^2 \sigma (pp\rightarrow \gamma^\prime pp)}{\dd z \dd k^2_\perp}\right]_{BB} = w_{\gamma'p}(z, k_\perp^2)\sigma_{pp}(\overline{s}),
\end{equation}
where the splitting function
\begin{multline} \label{eq:w}
				w_{\gamma'p}(z, k_\perp^2)=\frac{\epsilon^2 \alpha}{2\pi H} \left[\frac{1+(1-z)^2}{z}-2z(1-z)\left(\frac{2M^2+m^2_{\gamma'}}{H}-z^2\frac{2M^4}{H^2}\right)+
				\right.\\\left. + 2z(1-z)(1+(1-z)^2)\frac{M^2 m_{\gamma'}^2}{H^2}+2z(1-z)^2\frac{m^4_{\gamma'}}{H^2}\right],
\end{multline}
the frequently used kinematic combination $H\equiv k_\perp^2+(1-z)m_{\gamma'}^2+z^2M^2$ includes the proton mass $M$ and $\sigma_{pp}(\overline{s})$ is the elastic $pp$-scattering cross section taken at the square of protons center-of-mass energy. It was shown in \cite{Foroughi-Abari:2021zbm} that the BB approach does not agree with another available estimate for \textit{inelastic} bremsstrahlung. The result obtained in this approach for \textit{elastic} bremsstrahlung was also criticized in \cite{Gorbunov:2023jnx}. Despite this, it is actively used for estimating experimental sensitivities of future and proposed fixed-target experiments \cite{SHiP:2020vbd,Breitbach:2021gvv,Araki:2023xgb}.

The main thing that WW and BB approximations have in common is that the minus square of momentum transferred between protons $t$ or the momentum $q$ itself is set to 0. One can think about the validity of such approach for the composite particle such as proton. As it is shown in figure 1 in \cite{Gorbunov:2023jnx}, the maximum of the differential flux of hypothetical bosons that mediate elastic $pp$-interaction lies at $\sqrt{t}\simeq 0.35\text{ GeV}$. In contrast to electron bremsstrahlung, where the maximum of analogous virtual photon flux lies at $\sqrt{t_\text{min}} \ll \Lambda_\text{QCD}$ \cite{Kim:1973he}, this scale cannot be treated as negligible for the proton bremsstrahlung. Thus in \cite{Gorbunov:2023jnx} we considered the elastic $2\rightarrow 3$ bremsstrahlung taking into account non-zero momentum transfer $q$. We briefly summarize our main analytical results in the next section. 

\section{Bremsstrahlung with non-zero momentum transfer} \label{sec:calculation}
The full calculation of the elastic bremsstrahlung with non-zero momentum transfer is presented in \cite{Gorbunov:2023jnx}. There we considered the sum of two Feynman diagrams with the dark photon emission in the initial state and in the final state (for the last one see figure~\ref{fig:BBscheme}). In the lab frame the momenta marked in figure~\ref{fig:BBscheme} are as follows
\begin{align}
		P_i^\mu &=\{M, 0, 0, 0\}, \\
            p^\mu &=\{P+\frac{M^2}{2P}, 0, 0, P\}, \\
		k^\mu &=\{zP+\frac{m^2_{\gamma '}+k^2_\perp}{2zP}, k_x, k_y, zP\}, \\
		p'^\mu &=\{p'_0, -k_x-q_x, -k_y-q_y, P(1-z)-q_z\}, \\
		q^\mu &=\{q_0, q_x, q_y, q_z\},
\end{align}
where
\begin{equation}
p'_0=P(1-z)+\frac{M^2+k^2_\perp}{2P\left(1-z\right)}+\frac{k_xq_x+k_yq_y}{P\left(1-z\right)}-q_z+\frac{q^2_\perp}{2P\left(1-z\right)},
\end{equation}
$q^\mu$ is the hypothetical boson momentum going down from incident proton $p$ to the target proton $P_i$ in terms of figure~\ref{fig:BBscheme} and we assume that longitudinal components of momenta $\vec{p}$, $\vec{k}$ and $\vec{p^\prime}$ are much greater than their transverse components and the masses of these particles.
We consider the incident proton as a Dirac fermion as a whole and model the $pp$-interaction using photon-like propagator and the data on elastic $pp$-scattering.

The final result for differential cross section of proton bremsstrahlung takes the form
\begin{multline} \label{eq:bremcrsec}
\frac{\dd[2]{\sigma}}{\dd{k_\perp^2} \dd{z}} = \frac{\epsilon^2 \alpha}{32\left(2\pi \right)^2 z P \tilde{S}^2  \sqrt{P^2(1-z)^2+k_\perp^2}} \int\displaylimits_{t_\text{min}}^{t_\text{max}} \dd t \abs{T_+ + T_+^c}^2 \times \left[-b_0-\right. \\ \left. -\frac{b_1 t}{2M} + b_4 t+\left(b_2+\frac{b_5 t}{2 M}\right)\frac{k^2_\perp|\vec{q}|\cos \hat{\theta}_q}{\sqrt{P^2(1-z)^2 + k_\perp^2}}-\frac{b_3 k_\perp^2}{P^2(1-z)^2+k_\perp^2}\times\right.\\\left.\times\left(\frac{t}{2}\left(\frac{t}{4M^2}+1\right)P^2(1-z)^2+|\vec{q}|^2\cos^2\hat{\theta}_q\left(k_\perp^2 - \frac{P^2}{2}(1-z)^2\right)\right)\right],
\end{multline}
where $\tilde{S}^2\equiv\left(\bar{s}-2M^2\right)^2+q^2\left(\bar{s}-M^2\right)$, the limits of integration are $\sqrt{t_\text{min}}={H}/\left(2z(1-z)P\right)$ and $\sqrt{t_\text{max}}\sim m_{\gamma '}$, $T_+\left(\bar{s}, q^2\right)$ and $T_+^c\left(\bar{s}, q^2\right)$ are the fit functions for the differential cross section of elastic proton scattering defined as in \cite{ParticleDataGroup:2016lqr}, the coefficients $b_0,\dots,b_5$ are
\begin{align}
b_0&\equiv\frac{1+\left(1-z\right)^2}{1-z}, \label{eq:b0}\\
b_1&\equiv\frac{4zP}{H}\left(1+\left(1-z\right)^2\right), \\
b_2&\equiv-\frac{8\left(1-z\right)z^2}{H^3}\left(2M^2+m^2_{\gamma '}\right)\left(m^2_{\gamma '}+k^2_\perp-z^2M^2\right)-\\&\quad\, -\frac{2z\left(-1+\left(1-z\right)^2\right)}{H\left(1-z\right)},\\
b_3&\equiv\frac{16\left(1-z\right)^2 z^4}{H^4}\left(2M^2+m^2_{\gamma '}\right)P^2+\frac{z^2\left(-12+4\left(1-z\right)^2\right)}{H^2\left(1-z\right)}+\\&\quad\, +\frac{32}{H^4}\left(1-z\right)z^3\left(2M^2+m^2_{\gamma '}\right)\left(m^2_{\gamma '}+k^2_\perp-z^2M^2\right),  \label{eq:b3}\\
b_4&\equiv\frac{4\left(1-z\right)z^2P^2}{H^2}\left(1+\left(1-z\right)^2\right), \label{eq:b4}\\
b_5&\equiv-\frac{8z^2P}{H^2}\left(\left(1-z\right)^2+\frac{z^2}{H}\left(1-z\right)\left(2M^2+m^2_{\gamma'}\right)\right)
\end{align}
and the cosine of the angle between $\vec{q}$ and $\vec{p}-\vec{k}$ is fixed by the $\delta$-function
\begin{equation}
\cos \hat{\theta}_q = \frac{1}{2|\vec{p}-\vec{k}||\vec{q}|}\left(\frac{H}{z}+\frac{t}{M}\left(P(1-z)+M+\frac{M^2}{2P}-\frac{m^2_{\gamma '}+k^2_{\perp}}{2zP}\right)\right).
\end{equation}

\section{Numerical results} \label{sec:numerical}
\begin{figure}[t]
	\begin{center}
		\begin{subfigure}{0.48\textwidth}
			\centering
                \includegraphics[width=\textwidth]{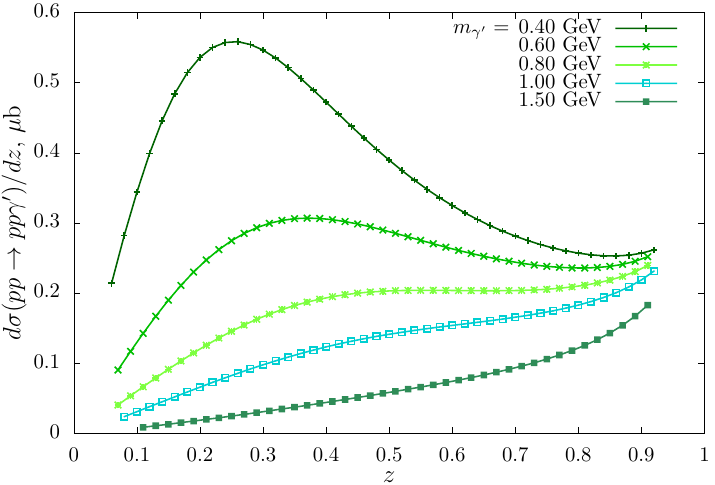}
			\caption{}
			\label{fig:crSec-z}
		\end{subfigure}
		\hfill
		\begin{subfigure}{0.48\textwidth}
			\centering
			\includegraphics[width=\textwidth]{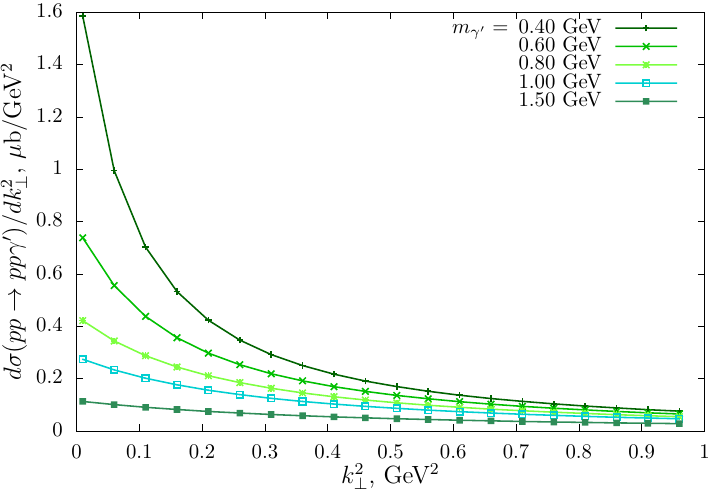}
			\caption{}
			\label{fig:crSec-ktr}
		\end{subfigure}
	\end{center}
	\caption{Differential cross section of elastic proton bremsstrahlung \eqref{eq:bremcrsec} depending on (a) the ratio of $z$-component of dark photon momentum to the momentum of incident proton, (b) the square of transverse momentum of a dark photon, both in the lab frame for different values of dark photon mass listed in the legend and the incident proton momentum $P=120\text{ GeV}$.}
\end{figure}

In this section we present the numerical results obtained for the incident proton beam with $P=120\text{ GeV}$ corresponding to the DUNE experiment. Figures~\ref{fig:crSec-z},\,\ref{fig:crSec-ktr}
show the differential cross section \eqref{eq:bremcrsec} integrated over the square of dark photon transverse momentum $k_\perp^2$ in the region 0.01-1\,GeV$^2$ ($\dd \sigma / \dd z$ in figure \ref{fig:crSec-z}) or over the momentum ratio $z$ in the region where $zP$ and $(1-z)P$ are much greater than transverse dark photon momentum and particle masses $m_{\gamma^\prime}$, $M$ ($\dd \sigma / \dd k^2_\perp$ in figure \ref{fig:crSec-ktr}). One can see that depending on the dark photon mass (we varied it in range 0.4-1.5\,GeV) the differential cross section $\dd \sigma / \dd z$ can behave in qualitatively different ways: either reaching the maximum at the intermediate values of $z$ or at $z\rightarrow 1$. Another finding is that the differential cross section $\dd \sigma / \dd k^2_\perp$ quickly decreases with the growth of transverse momentum squared $k_\perp^2$ which is important for understanding which part of produced dark photons will reach a detector.

\begin{figure}[t]
	\begin{center}
		\begin{subfigure}{0.48\textwidth}
			\centering
                \includegraphics[width=\textwidth]{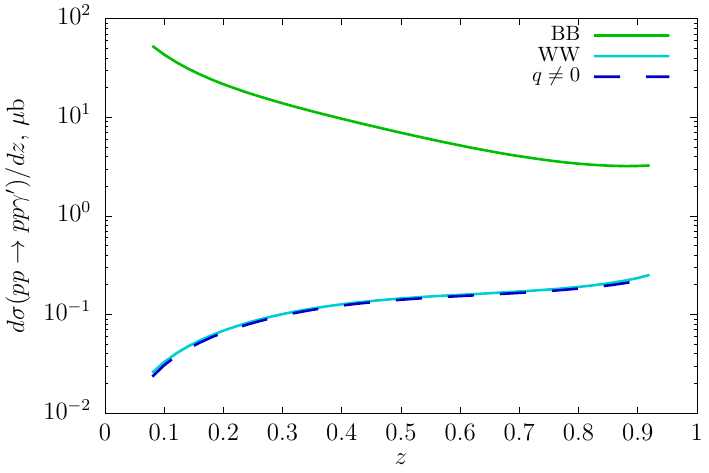}
			\caption{}
			\label{fig:crSec-z-comp}
		\end{subfigure}
		\hfill
		\begin{subfigure}{0.48\textwidth}
			\centering
			\includegraphics[width=\textwidth]{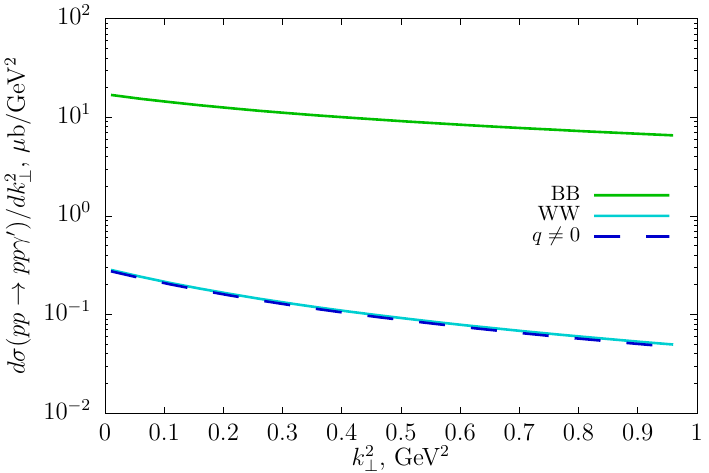}
			\caption{}
			\label{fig:crSec-ktrsq-comp}
		\end{subfigure}
	\end{center}
	\caption{The comparison of differential cross section with non-zero momentum transfer \eqref{eq:bremcrsec} \cite{Gorbunov:2023jnx} (dashed line) with the generalized WW approximation (light blue line) \cite{Kim:1973he} and the result of BB (green line) \cite{Blumlein:2013cua} depending on (a) the ratio of $z$-component of dark photon momentum to the momentum of incident proton, (b) the square of transverse momentum of a dark photon, both in the lab frame for the incident proton momentum $P=120\text{ GeV}$ and the dark photon mass $m_{\gamma^\prime}=1\text{ GeV}$.}
\end{figure}

In figures~\ref{fig:crSec-z-comp},\,\ref{fig:crSec-ktrsq-comp}
we compare the differential cross section of elastic proton bremsstrahlung obtained in our work \cite{Gorbunov:2023jnx} with the result in the generalized Weizsacker-Williams approximation \cite{Kim:1973he} and the one of Blumlein and Brunner \cite{Blumlein:2013cua}. In the latter case we multiply the splitting function $w_{\gamma'p}(z, k_\perp^2)$ \eqref{eq:w} by the elastic $pp$-cross section. The growth of BB cross section with $z\rightarrow 0$ is tightly connected with the structure of the splitting function \eqref{eq:w} containing the terms proportional to $1/z$. One can observe the good agreement (within 3-9\,\%) between our result taking into account $q\neq 0$ and the WW approximation. On the contrary, the result of BB, as was also earlier noticed in \cite{Foroughi-Abari:2021zbm} for \textit{inelastic} bremsstrahlung, significantly differs from other available estimates.

In order to justify the observed agreement, one can study the dependence of the differential cross section $\dd^3 \sigma / \dd z \dd k^2_\perp \dd t$ on $t$ for the generalized WW approximation and for our refined answer. The former is analogous to the differential flux of hypothetical bosons in the elastic $pp$-scattering $\dd \chi_b / \dd t$ depicted in figure 1 in \cite{Gorbunov:2023jnx}. They have the same shape, and the maxima of these distributions for proton elastic bremsstrahlung lie at close $\sqrt{t}\sim \Lambda_\text{QCD}$. The generalized WW approximation can be obtained from our answer by neglecting the terms with coefficients $b_1$, $b_2$, $b_5$ and partly $b_3$ in \eqref{eq:bremcrsec} and using that $\sqrt{t_\text{min}}\ll M$, $k^2_\perp \ll P^2(1-z)^2$ and $t \ll 4M^2$ for minus square of transferred momentum $t$ with $\dd \chi_b / \dd t$ that is significantly different from 0. It is important to note that the last condition is stronger than the naive one $\sqrt{t}\lesssim M$, coming from the comparison of typical scales in this problem mentioned in the end of section \ref{sec:approximations}. 

The generalized WW approximation, originally formulated for one-photon exchange processes with electrons, explicitly utilized the fact that the maximal contribution to the differential cross section is at $\sqrt{t}\sim \sqrt{t_\text{min}}$ due to the singular behaviour of the matrix element at small $t$. Opposite to this, we see that for proton bremsstrahlung the differential cross section maximum is reached at the momentum transfer $\sqrt{t}\sim \Lambda_\text{QCD}$ that is small compared to the momentum $P$, but is still non-negligible and that there is no singularity at $\sqrt{t}/M\rightarrow0$. Generally speaking, since small momentum transfers can be associated with interaction at large distances, this is the illustration of the fact that the electromagnetic interaction is long-range (photon exchange in electron bremsstrahlung), while the strong interaction playing the key role in proton bremsstrahlung has a finite range. Despite this, the coincidence of distribution shapes and the conditions mentioned earlier like $t\ll 4M^2$ allow one to successfully use the WW approximation. 

\section{Conclusions} \label{sec:conclusions}
To sum up, we have obtained the elastic bremsstrahlung cross section considering the non-zero momentum transfer between incident and target protons. Our result agrees well with the generalized WW approximation \cite{Kim:1973he}. Thus the generalized WW approximation is applicable to protons and can give reliable results.

\section*{Acknowledgements}
We thank D.~Gorbunov for helpful advice, continuous support and patience. We are also grateful to O.~Teryaev and I.~Timiryasov for valuable comments and discussion. The work is partially supported by the Russian Science Foundation RSF grant 21-12-00379 (analytical calculation) and the grant of  Foundation for the Advancement of Theoretical Physics and Mathematics “BASIS” no. 21-2-10-37-1 (numerical results).

\end{document}